%% file: main.tex
\documentclass[
    twocolumn,
	prd,
	amssymb,
	preprintnumbers,superscriptaddress,
	nofootinbib]{revtex4-1}

\pdfoutput=1

\usepackage{graphicx}
\usepackage{enumitem}
\usepackage{latexsym}
\usepackage{amsfonts}
\usepackage{amssymb}
\usepackage{xcolor}
\usepackage[export]{adjustbox}
\usepackage{amsmath}
\usepackage{slashed}
\usepackage{dcolumn}
\usepackage{verbatim}
\usepackage{float}
\usepackage{multirow}
\usepackage{xspace}
\usepackage{hyperref}
\usepackage[normalem]{ulem}

\input{universalnewcommands.tex}

\newcommand{\dd}{\mathrm{d}}

\newcommand{\msun}{{\rm M}_\odot}
\newcommand{\rC}{{\rm C}} \newcommand{\rO}{{\rm O}}

\allowdisplaybreaks

\begin{document}

\title{Modified Gravity and the Black Hole Mass Gap}
\author{Maria C. Straight} \email{mstraight21@my.whitworth.edu }
\affiliation{Department of Engineering and Physics, Whitworth University, 300 W. Hawthorne Rd., Spokane, WA 99251}
\author{Jeremy Sakstein} \email{sakstein@hawaii.edu}
\affiliation{Department of Physics \& Astronomy, University of Hawai'i, Watanabe Hall, 2505 Correa Road, Honolulu, HI, 96822, USA}
\author{Eric J. Baxter}
\email{ebax@hawaii.edu}
\affiliation{Institute for Astronomy, University of Hawai'i, 2680 Woodlawn Drive, Honolulu, HI 96822}

\date{\today}

\begin{abstract}
We pioneer the black hole mass gap as a powerful new tool for constraining modified gravity theories. These theories predict fifth forces that alter the structure and evolution of population-III stars, exacerbating the pair-instability. This results in the formation of lighter astrophysical black holes and lowers both the upper and lower edges of the mass gap. These effects are explored using detailed numerical simulations to derive quantitative predictions that can be used as theoretical inputs for Bayesian data analysis. We discuss detection strategies in light of current and upcoming data as well as complications that may arise due to environmental screening. To demonstrate the constraining power of the mass gap, we present a novel test of the strong equivalence principle where we apply our results to an analysis of the first ten LIGO/Virgo binary black hole merger events to obtain a {$7\%$} bound on the relative difference between the gravitational constant experienced by baryonic matter, and that experienced by black holes, $\Delta G/G$. The recent GW190521 event resulting from two black holes with masses in the canonical mass gap can be explained by modified gravity if the event originated from an unscreened galaxy where the strength of gravity is either enhanced or reduced by $\sim30\%$ relative to its strength in the solar system. 
\end{abstract}

\maketitle

\section{Introduction}
\label{s.Introduction}

The advent of gravitational wave astronomy has opened a new window to the Universe and furnishes us with new tools for testing the laws of nature. Indeed, the observation of gravitational waves by the LIGO/Virgo gravitational wave interferometers has already allowed for consistency tests of general relativity (GR) on new scales, and at extreme space-time curvatures \cite{TheLIGOScientific:2016src,Abbott:2018lct,LIGOScientific:2019fpa}. The majority of observations to date have been of merging black holes. These are useful tools to test small scale (ultra-violet) modifications of gravity (e.g. \cite{Berti2015-testGR}) that could modify the inspiral and merger dynamics but are less useful for constraining long distance (infra-red) modifications of gravity that could drive the acceleration of the cosmic expansion (i.e. act as dark energy) \cite{Copeland2006-DarkEnergy,Koyama2016-CosmologicalTests,Joyce2015-BeyondStandard,Burrage2018-ChameleonGravity,Sakstein2018-testScreened,Baker2019-GravityTest} or resolve the Hubble tension \cite{Sakstein2019-Screened,Desmond2019-resolveHT,Desmond2020-Screened}. The underlying reason for this is that these theories typically introduce new light degrees of freedom that couple to gravity, most commonly scalars. Such theories are typically subject to powerful no-hair theorems precluding any modifications of the black hole solutions \cite{Bekenstein:1995un,Mazur2000-nohair,Hui:2012qt} (see \cite{Sotiriou:2013qea,Babichev:2016rlq,Silva:2017uqg,Antoniou:2017acq,Noller:2019chl} for some notable exceptions). The torrent of binary black hole mergers (BBHMs) expected in the upcoming LIGO/Virgo data release and beyond provides strong motivation for the development of novel probes of infra-red modifications of gravity that utilize such observations. In this work, we explore one such probe: the black hole mass gap (BHMG). 

The BHMG refers to the predicted absence of astrophysical black holes with masses in the range $\sim 50--120~{\rm M}_{\odot}$ (there is some uncertainty due to environment and stellar uncertainties). The origin of the BHMG is the pair-instability \citep{Rakavy1967-StellarInstabilities,Fraley1968-PPIexplosions,Bond1984-VMO}. The core temperatures and densities of massive stars are sufficient for the thermal production of electron-positron pairs from the plasma. These act to destabilize the star by reducing the pressure and causing a gravitational contraction. The resultant increase in temperature can ignite oxygen explosively. What happens next depends on the star's mass. Stars with initial helium core masses between $M\sim 40$--$60$ ${\rm M}_\odot$ (assuming metallicity $Z=10^{-5}$) experience a series of nuclear flashes that drive strong pulsations and cause severe mass ejections referred to as a pulsational pair-instability supernova (PPISN). These pulsations are not energetic enough to disrupt the entire star and {the star} ultimately returns to hydrostatic equilibrium before collapsing to form a black hole. The resultant black hole is significantly less massive than the original star. For stars with initial helium core masses between $\sim 60-130$ M$_\odot$ (for metallicity $Z=10^{-5}$), the explosion is so violent that the entire star becomes unbound and no black hole is left behind. This process is referred to as a pair-instability supernova (PISN). The heaviest black hole that can be formed as a result of the competition between the PPISN and the PISN defines the lower edge of the BHMG. At higher masses ($M\sim120{\rm M}_\odot$ for $Z=10^{-5}$) the PISN is quenched because some of the energy in the contraction is utilized to photodisintegrate the heavy elements instead of raising the temperature to ignite the oxygen, and black holes can once again be formed after collapse of the star. The black hole formed from the lightest star that does not experience a PISN defines the upper edge of the BHMG.\footnote{{We note that there is also evidence for a second black hole mass gap in the range of 2 to $5\,M_{\odot}$, between the maximum neutron star mass and the lowest black hole mass (e.g. \cite{Farr:2011,Fishbach:2020ryj}).  We do not consider the impact of modified gravity on this lower mass gap.  As we discuss in \S\ref{s.data}, the precise value of the lowest black hole mass has a small impact on our results.}}

{Current observations by LIGO provide tentative evidence for the existence of a BHMG \cite{Fishbach2017-BigBlackHoles,Abbott2019-LIGOBBH}, with future upgrades to LIGO expected to improve upon these constraints \cite{Ezquiaga2020-jumpthegap}.} Using the first four LIGO detections, \citet{Fishbach2017-BigBlackHoles} constrained the lower edge of the BHMG to be at about 41 $\rm{M}_\odot$, while the largest black hole observed by LIGO/Virgo to date has a mass $M_{\rm g} = 50.2^{+16.2}_{-10.2}\ \rm{M}_{\odot}$
\citep{Abbott2019-LIGO} {(consistent at $\sim 1\sigma$ with the inferred gap location)}. As LIGO has upgraded the sensitivity of its detectors and expects a significant increase in the number of detections in the coming O3 run and beyond, now is an opportune time to consider how to use this upcoming data to test new physics. Indeed, references \cite{Croon:2020ehi} and \cite{Croon:2020oga} have recently investigated the effects of light particle emission on the black hole mass gap, with promising results. {LIGO/Virgo recently announced the detection of a binary black hole merger event, GW190521, with component masses $m_1 =85^{+21}_{-14}{\rm M}_\odot$ and $m_2=66^{+17}_{-18}{\rm M}_\odot$, both of which are located in the GR mass gap  \cite{PhysRevLett.125.101102,Abbott_2020}. We will briefly comment on this detection and the modified gravity scenarios under which they could have formed.}

The black holes detected by LIGO/Virgo may be insensitive to long distance modifications of GR {(because of no-hair theorems)}, but their progenitors are not. Indeed, a ubiquitous feature of {theories of} light scalars coupled to gravity is screened fifth forces \cite{Burrage2018-ChameleonGravity,Sakstein2017-TestsofGravity,Sakstein2018-testScreened}, which can alter the structure and evolution of stars \cite{Davis:2011qf,Jain:2012tn,Sakstein:2013pda,Sakstein:2014nfa,Koyama:2015oma,Saito:2015fza,Sakstein:2015zoa,Sakstein:2015oqa,Sakstein:2015aac,Jain:2015edg,Babichev:2016jom,Sakstein:2016lyj,Saltas:2019ius,Olmo:2019flu,Sakstein:2019qgn, Wojnar:2020txr}. Screening refers to the strong environmental-dependence of the modifications of GR that is necessary for the theories to simultaneously explain the large-scale mysteries such as dark energy and the Hubble tension and satisfy solar system and laboratory tests of gravity \cite{Will2014-GRexperiment,Sakstein:2017xjx,Burrage2018-ChameleonGravity}. These features make them leading science targets for upcoming missions such as Euclid \cite{Amendola:2016saw}, and for this reason they will be the focus of the work presented here. 

Screening mechanisms fall into three categories. Thin-shell screening such as the chameleon \citep{Khoury2004-Chameleon}, symmetron \cite{Hinterbichler:2010es}, and dilaton \cite{Brax:2010gi} mechanisms screen fifth forces by suppressing the scalar charge of individual objects so that they do not respond to external fifth forces. Kinetic screening models such as Vainshtein screening \cite{Vainshtein:1972sx,Nicolis2009-Vainshtein} and K-mouflage \cite{Babichev:2009ee} instead act to suppress the force fields themselves. In a third class of theories, an interaction between dark matter and baryons causes the value of Newton's constant to become dependent on the local dark matter density \cite{Sakstein2019-Screened}, altering its values in regions less dense than the solar neighborhood. {Independent of the specific mechanism, in an unscreened environment the effective value of the gravitational constant would be modified to}
\begin{equation}\label{eq:dgdef}
    G=\left(1+\frac{\Delta G}{G_{\rm N}}\right)G_{\rm N},
\end{equation}
where $G_{\rm N}$ is the value of Newton's constant measured in the solar system \cite{Desmond2019-resolveHT} i.e. $\Delta G=0$ corresponds to screened environments. The screening nature is encapsulated because $G$ is environment-dependent.\footnote{The manner of this dependence depends on the theory at hand. See \cite{Desmond2019-resolveHT} for a discussion of how $\Delta G/G_{\rm N}$ correlates with different astrophysical environments in different theories.} This implies that the progenitors of the black holes observed by LIGO/Virgo {may have}  evolved under a different value of $G$. The parameterization in equation \eqref{eq:dgdef} is exact for DHOST theories \cite{CrisostomiKoyama2018,Dima:2017pwp,Langlois:2017dyl,Crisostomi:2019yfo}, which are the leading candidate dark energy theories after some simpler ones were excluded by multimessenger observations of merging neutron stars (GW170817) \cite{Sakstein:2017xjx,Baker:2017hug,Ezquiaga:2017ekz,Creminelli:2017sry}. 
Similarly, the parameterization is exact for the dark-matter--baryon screening mechanism \cite{Sakstein2019-Screened}, which can both account for dark energy \cite{Berezhiani:2016dne} and resolve the Hubble tension \cite{Desmond2019-resolveHT,Desmond2020-Screened}. Other screening mechanisms either have additional parameters that control the efficiency of screening as a function of the environmental variables, or $\Delta G/G_{\rm N}$ is radially-dependent throughout the star. {In the latter case, one should think of $\Delta G$ as an average over the object.}

In this work we will be theory-agnostic and investigate the effects of changing $G$ on the location of the black hole mass gap. We postpone the more arduous task of correlating each BBHM event with its environment for future work, although we discuss potential detection strategies in our conclusions. We numerically simulate the evolution of population-III stars from zero age helium burning (ZAHB) to either core collapse or PISN, finding that increasing $G$ results in a stronger instability. The effects of this are twofold. First, the mass loss during pulsations is increased, resulting in lighter black holes being formed. Second, lighter objects that would have resulted in a PPISN instead experience a full PISN, removing black holes from the distribution. {Using current observations from LIGO, we set limits on departures of $G$ from $G_{\rm N}$, constraining $\Delta G/G_{\rm N}$ to 7\% precision. This proof of principle analysis demonstrates the exciting potential of the BHMG as a probe of modified gravity; as discussed in \S\ref{s.data} and Appendix~\S\ref{app:toymodel}, constraints are expected to improve quickly with larger samples of BBHM events.} We assume throughout that modified gravity impacts the progenitor stars of the BBH system, but that once the black holes have formed, their subsequent interaction is governed by GR {i.e. $\Delta G/G_{\rm N}=0$ for black holes}; this assumption is well motivated by the no-hair theorems, as we discuss in \S\ref{s.waveform}, {{and is tantamount to assuming that the waveforms are identical to GR, as are the inferred black hole masses}}. {For this reason, our bound does not constitute a measurement of Newton's constant but rather a novel test of the strong equivalence principle between black holes and baryonic matter, which is a unique feature of GR.}

This paper is organized as follows. In section \ref{s.BHMG} we provide an introduction to the black hole mass gap for the benefit of the unfamiliar reader and discuss the main sources of uncertainty in its precise location. In section~\ref{s.modeling}, we present our numerical methods for modeling stellar evolution. We discuss the effects of changing $G$ on individual stars, and on the location of the BHMG from a grid of stars with different masses in section~\ref{s.results}. In section \ref{s.data} we present our statistical analysis of the first ten LIGO/Virgo binary black hole events, finding a constraint on $\Delta G/G_{\rm N}$. We conclude in section~\ref{s.concs} by discussing possible detection strategies, potential complications due to environmental screening, and future work.

\section{Physics of the Black Hole Mass Gap}
\label{s.BHMG}

In this section we briefly introduce the physics of the BHMG and discuss various sources of uncertainties due to environment, stellar modeling, and input physics.

We begin with the origin of the instability. The pair-instability is due to the production of electron-positron pairs from the thermal plasma once the core density and temperature are sufficiently high. The threshold photon energy for producing such pairs is $E_{\gamma\gamma}=2m_e$, corresponding to temperatures of  $10^{10}$K, but the onset of the instability occurs for  temperatures of order $8.5\times10^{8}$K. The reason for this is the large number of photons present in the stellar interior, which partially compensates for the exponential suppression of the Bose-Einstein distribution at high photon energies. The electron-positron pairs are produced with non-relativistic velocities, which has the effect of lowering the equation of state (EOS), or first adiabatic index,
\begin{equation}
    \Gamma_1\equiv \left(\frac{\partial\log P}{\partial\log\rho}\right)_s.
\end{equation}
The stars are radiation dominated so in the absence of pair-production the EOS is $\Gamma_1\approx4/3$. This is the threshold below which stars are unstable (see e.g. \cite{Cox1968-stellarstructure,Kippenhahn2012-Stellar}) and the reduction due to pair-production therefore destabilizes the star. In essence, the $e^+e^-$ pairs contribute to the density but not the pressure, so their production robs the star of its pressure support, resulting in gravitational contraction. As the star contracts, the resultant increase in temperature and density does not raise the pressure to counteract the contraction but instead causes an increase in the production rate of $e^+e^-$ pairs. The situation is only reversed when the temperature and density are high enough to ignite oxygen explosively.

\begin{figure}
    \centering
    \includegraphics[width=3.25in]{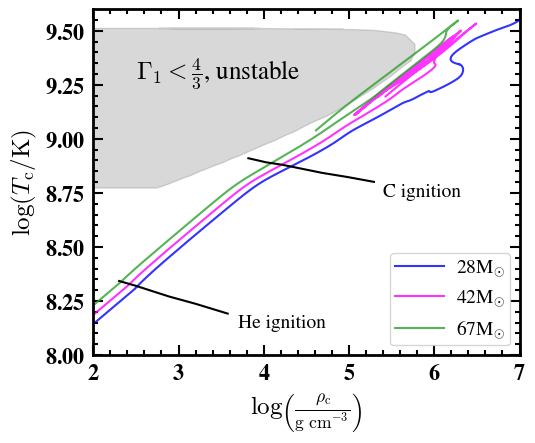}
    \caption{The evolution of the central temperature versus the central density of population-III stars with initial masses $M_{\rm in} = 28\msun$ (blue), $42\msun$ (magenta) and $67\msun$ (green) with initial metallicity  $Z=10^{-5}$. The gray region indicates the area where the pair-instability occurs, and the black lines indicate the onset of helium and carbon burning.}
    \label{fig:Trho_GR}
\end{figure}

The instability region in the $\log_{10}(T_c)$--$\log_{10}(\rho_c)$ plane is shown in Figure~\ref{fig:Trho_GR} along with some representative stellar tracks. Its shape can be understood as follows. The lower boundary corresponds to temperatures too low for $e^+e^-$ pairs to be produced in sufficient quantities to lower the EOS below $4/3$. As the temperature increases, the $e^+e^-$ pairs are produced with higher energies, and the upper edge corresponds to temperatures where they are produced with relativistic velocities. In this case the electron-positron EOS is $\Gamma_1\sim 4/3$ so the $e^+e^-$ pairs do contribute to the pressure and there is no instability. The right hand edge corresponds to densities where the gas pressure of the ions in the star is not negligible. These are non-relativistic so have EOS $\Gamma_1\sim 5/3$ and thus stabilize the star.

\begin{figure}
    \centering
    \includegraphics[width=3.25in]{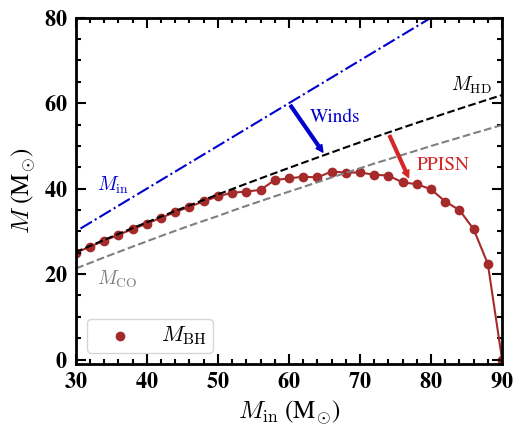}
    \caption{Various masses over the course of stellar evolution as a function of the initial (ZAHB) mass for population-III stars of initial metallicity $Z_\odot/10$. Stars with the initial masses shown by the blue dot-dashed line lose mass due to stellar winds. The black dashed line shows the star masses at helium depletion, and the gray dashed line shows the CO core masses at helium depletion. The red dots show the final black hole masses formed for each star after more mass has been lost due to pulsations. At higher masses, a PISN occurs and no black hole is formed.
    }
    \label{fig:MvsMin}
\end{figure}

The ultimate fate of the star depends on its mass, which determines the nature of oxygen ignition after the initial contraction. Lower mass stars (e.g. the 28${\rm M}_\odot$ track in Fig.~\ref{fig:Trho_GR}) miss the instability region entirely. In this case, heavy element fusion to iron proceeds non-explosively and the star ultimately undergoes core collapse to form a black hole of mass similar to its initial mass (minus a small amount of mass lost due to stellar winds). Intermediate mass stars (e.g. the 42${\rm M}_\odot$ track in Fig.~\ref{fig:Trho_GR}) graze the instability region and experience the contraction. After oxygen is ignited, the stars undergo a series of pulsations (a pulsational pair instability supernova) where large amounts of mass are shed. The star ultimately returns to hydrostatic equilibrium and collapses to form a black hole far lighter than the star's initial mass. Heavier stars (e.g. the 67${\rm M}_\odot$ track in Fig.~\ref{fig:Trho_GR}) experience the full instability and the explosive oxygen burning results in a thermonuclear explosion that unbinds the entire star (a pair instability supernova) and leaves no black hole remnant. In very massive stars, the PISN is quenched because some of the energy in the contraction is used to photodisintegrate heavy elements rather than to ignite oxygen. These stars do leave behind black hole remnants. The black hole mass gap is the unpopulated region formed from the interplay of these processes. The lower edge corresponds to the highest mass black hole that can be formed before the PPISN acts to significantly reduce the mass of higher mass progenitor stars (the turnover in Fig.~\ref{fig:MvsMin}). The upper edge corresponds to the lowest mass black hole that is formed as a result of the quenching of the PISN. 

An example of the black hole mass distribution as a function of initial (ZAHB) mass below the lower edge of the BHMG is shown in Fig.~\ref{fig:MvsMin}. Its shape can be understood as follows. During core helium burning, stars lose mass to stellar winds. For stars that avoid the instability region, this is the only source of mass loss so the final black hole mass is given by the mass at helium depletion i.e. the mass remaining after wind-losses have terminated. Stars that graze the instability region experience further mass loss due to the PPISN and so form black holes with masses $M_{\rm BH}<M_{\rm HD}$. As the initial mass is increased and the pulsations become increasingly violent the mass loss is significant enough that the distribution turns over and the black holes formed are lighter than those formed from lighter progenitors. Eventually, the stars undergo a PISN and no black holes are formed, corresponding to the steep fall to zero at high initial masses.

The precise location of the lower edge of the mass gap is subject to several sources of uncertainty due to environment (metallicity, binarity), stellar modeling (wind loss, rotation, mixing length, numerical resolution etc.), and uncertainties on the input physics (nuclear reaction rates, neutrino loss rates etc.). Reference \cite{Farmer2019-BHmassgap} has studied these in detail (see also \cite{Farmer2020-COreactionrate} for the case of nuclear reaction rates, \cite{Marchant2018-PPISNbinaries} for binarity, \cite{Marchant:2020haw} for rotation, and \cite{Renzo:2020rzx} for time-dependent convection). We summarize the most significant uncertainties below.
\begin{itemize}
    \item \textbf{Metallicity:} The rate of wind loss is proportional to $Z^{0.85}$ \cite{Vink2001-massloss,Brott2011-MainSequence} so more metal rich objects lose more mass to stellar winds. This results in a spread of $\sim 3{\rm M}_\odot$ \cite{Farmer2019-BHmassgap} in the location of the lower edge. Since we expect stars of all metallicities to exist this is less of an uncertainty for us and more of a feature, although it may become an important systematic once one tries to correlate specific gravitational wave observations with their host galaxies. For the purposes of this work, we are interested in a universal lower edge (rather than an environment-dependent one) so we will take $Z=10^{-5}$, which corresponds to the highest possible lower edge due to the minimum wind loss. The upper edge corresponds to $Z\sim10^{-3}$, so we will take $Z={\rm Z}_{\odot}/10=0.00142$ when studying the upper edge.
    \item \textbf{Wind Loss:} The wind loss prescription we use (see section \ref{s.modeling}) includes a free parameter $\eta$, which is an overall scaling factor for the rate. Varying this over the range $0.1\le\eta\le1.0$ results in a variation in the location of the lower edge of $\sim 3{\rm M}_\odot$ \cite{Farmer2019-BHmassgap}. We take $\eta=0.1$ corresponding to the fiducial value assumed by reference \cite{Farmer2019-BHmassgap}.
    \item \textbf{Nuclear reaction rates:} Uncertainties in the nuclear reaction rates propagate into the final black hole mass. The reason for this is that the strength, or even existence, of the PPISN is strongly dependent on the ratio of carbon to oxygen at the end of helium burning. A higher ratio of $^{12}\textrm{C}/^{16}\textrm{O}$ suppresses the PPISN (and PISN) \cite{Farmer2019-BHmassgap}. There are two competing rates that determine the $^{12}\textrm{C}/^{16}\textrm{O}$ ratio: the triple-$\alpha$ process, which converts $^4\textrm{He}$ to $^{12}\textrm{C}$, and the $^{12}\textrm{C}(\alpha,\gamma)^{16}\textrm{O}$ reaction, which converts $^{12}\textrm{C}$ to $^{16}\textrm{O}$. The latter is by far the most important reaction. Using the most up to date rates \cite{deBoer:2017ldl}, the uncertainty in the location of the lower edge is $M_{\rm lower}=51^{+0}_{-4}\rm{M}_\odot$. In this work we will use \acro{MESA}'s default reaction rate in order to allow for a direct comparison with previous works, so our lower edge (assuming GR) lies at $M_{\rm lower}\sim 47{\rm M}_\odot$, consistent with the results of \cite{Farmer2019-BHmassgap}. 
\end{itemize}
All other sources of uncertain input physics result in changes of $1{\rm M}_\odot$ or less in the location of the mass gap. The upper edge is subject to similar uncertainties.

\section{Stellar Modeling}
\label{s.modeling}

We numerically simulate helium cores from the ZAHB to either core collapse or PISN using the stellar structure code \acro{MESA} version 12778 \cite{Paxton2011-MESA,Paxton:2015jva,Paxton2018-MESA,2019ApJS..243...10P}, modified to change the value of $G$ for entire simulations. 

Our procedure closely follows the one described in \cite{Paxton2018-MESA,Marchant2018-PPISNbinaries,Farmer2019-BHmassgap} with one important exception during the PPISN phase. Individual pulses during a PPISN cause large fractions of the star's mass to be removed at high velocities, while the remaining material returns to hydrostatic equilibrium with a lower central temperature than before the pulse. \acro{MESA} cannot compute the long-term evolution of both the ejected mass and the bound core, so the unbound layers are removed from the model and a new stellar model is relaxed using the procedure described in Appendix B of \cite{Paxton2018-MESA} such that it has the same mass, entropy, and composition profiles as the material that was bound in the hydrodynamical model. Appendix C of \cite{Marchant2018-PPISNbinaries} explores how well the relaxation procedure reproduces the pre-relaxation model. As a starting point for the relaxation, MESA calls a pre-made zero age main sequence (ZAMS) model. The choice of initial model is arbitrary and the final state of the star post-relaxation is independent of the initial conditions. The ZAMS models come pre-packaged with MESA and, as such, assume GR so they are not in hydrostatic equilibrium for different values of $G$. This causes the relaxation process to fail. To remedy this, we have computed new ZAMS models for the values of $G$ we investigate in this work and call these during the relaxation process. All of the files necessary to reproduce our results are available at \url{https://zenodo.org/record/4037390}, including our routines to generate the modified ZAMS models.

We use the following prescriptions for stellar processes, which correspond to the fiducial choices of references \cite{Marchant2018-PPISNbinaries,Farmer2019-BHmassgap}. Mass loss due to stellar winds uses the prescription of \cite{Brott2011-MainSequence} and is proportional to $(Z/Z_\odot)^{0.85}$ \cite{Vink2001-massloss}, implying higher metallicity stars experience a greater mass loss. The free parameter $\eta$ that scales the overall mass loss rate is taken to be $\eta=0.1$ corresponding to the fiducial value of \cite{Farmer2019-BHmassgap}. The lower edge of the BHMG in GR is due to stars with metallicity $Z=10^{-5}$ so we use the same metallicity for our simulations of the lower edge. The location of the upper edge corresponds to higher metallicity stars so we take $Z={\rm Z}_\odot/10=0.00142$ when simulating upper edge progenitors. Convection is modeled using mixing length theory (MLT) \cite{Cox1968-stellarstructure} with efficiency parameter $\alpha_{\rm MLT}=2.0$ and semi-convection is modeled using the  prescription of \cite{Langer1985-semiconvection} with efficiency parameter $\alpha_{\rm sc}=1.0$. Convective overshooting is described by an exponential profile which has two free parameters: $f_0$, which sets the point inside the convective boundary where overshooting begins, and $f_{\rm ov}$, which sets the scale height of the overshoot. We set these parameters to the fiducial choices made by \cite{Farmer2019-BHmassgap} $f_0=0.005$ and $f_{\rm ov}=0.01$. The nuclear reaction rates are set to the \acro{MESA} defaults, which are a mixture of the {\tt NACRE} \cite{Angulo:1999zz} and {\tt REACLIB} \cite{2010ApJS..189..240C} rates. The other controls are set to the recommended values given in the \emph{test\_suite}~ {\tt ppisn} (which comes prepackaged with \acro{MESA}) with the exception of {\tt mesh\_delta\_coeff}, which determines the number of cells used by \acro{MESA}'s adaptive grid. We set this to $0.5$.

Quantities of interest are the mass at helium depletion, the carbon-oxygen (CO) core mass, and the black hole mass. These are defined as in \cite{Marchant2018-PPISNbinaries} and \cite{Farmer2019-BHmassgap} to allow for direct comparisons. In particular, helium depletion is defined as the time when the central helium mass fraction falls below $0.01$. The CO core mass is defined at this time as the innermost mass coordinate with a helium mass fraction greater than $0.01$. The black hole mass is calculated at core collapse as the mass of bound material. This is defined as the mass within the outermost coordinate in which the layer's velocity is less than the escape velocity $v_{\rm esc}=\sqrt{2GM/R}$ (note that this depends on $G$).  If all of the layers are expanding with a velocity $v>v_{\rm esc}$ during a pulse then then entire star has become unbound, signaling a PISN. No black hole is formed in this case.

\section{Effects of Modified Gravity on the Black Hole Mass Gap}
\label{s.results}
There are several competing effects of changing the strength of gravity on the evolution of population-III stars. We discuss these in detail below, exemplifying them by considering values of $G$ larger than $G_{\rm N}$ ($\Delta G/G_{\rm N}>0$) since the screening mechanisms we have in mind typically increase the strength of gravity. We will also study $\Delta G/G_{\rm N}<0$ when discussing the BH mass distribution since some theories (e.g. beyond Horndeski and subsets of DHOST) have this qualitative effect.

\begin{figure}[h]
    \centering
    \includegraphics[width=3.25in]{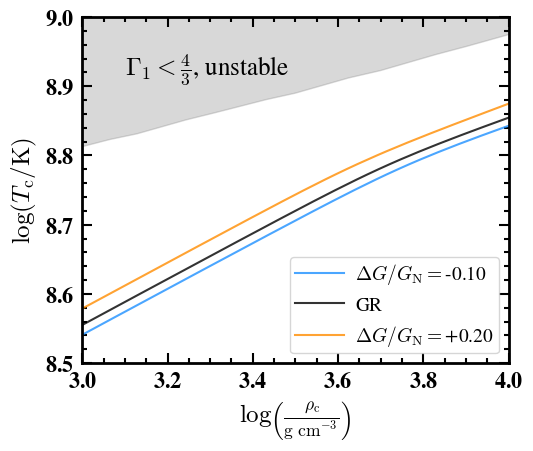}
    \caption{Stellar tracks in the $\log T_{\rm c}$--$\log{\rho_{\rm c}}$ plane for values of $G$ indicated in the figure. The tracks correspond to stars with metallicity $Z=10^{-5}$ and zero age helium branch mass 56$\msun$. The pair-instability region is indicated in gray.
    }
    \label{fig:Trho}
\end{figure}
\begin{figure}
    \centering
    \includegraphics[width=3.25in]{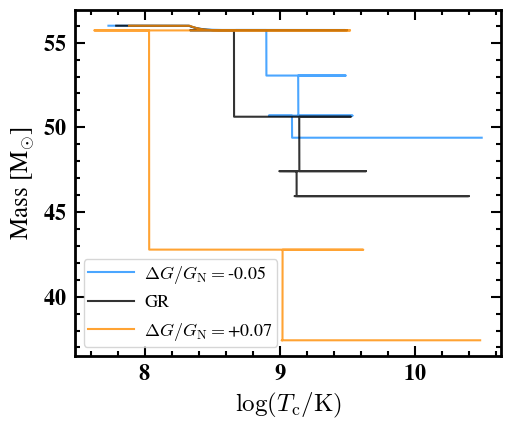}
    \caption{Mass versus central temperature (a proxy for time) for a star with initial mass $56{\rm M}_\odot$ and initial metallicity $Z=10^{-5}$ for GR and the values of $\Delta G/G_{\rm N}$ shown. The star evolves from low to high temperatures, but this trend is reversed during pulsations. The sudden vertical drops indicate a single pulse.
    }
    \label{fig:MassvsTc}
\end{figure}

The first effect is that the structure and evolution of the star is altered because the pressure support required to maintain hydrostatic equilibrium is increased. Indeed, consider the hydrostatic equilibrium equation
\begin{equation}
\frac{\dd P}{\dd r}= -\frac{GM(r)\rho(r)}{r^2}
\end{equation}
which must be satisfied for the pressure support to balance the inward gravitational force. Considering a radiation pressure-supported star with $P\propto T^4$, we can obtain scaling relations for a star of mass $M$ and core radius $R$ by setting $\dd/\dd r\sim R^{-1}$, $r\sim R$, $M(r)\sim M$, and $\rho(r)\sim M/R^3$ to find
\begin{equation}
    \log(T_c)=\frac{1}{3}\log(\rho_c)+\frac14\log(G)+\frac{1}{6}\log(M)+c,
\end{equation}
where $c$ is a constant that is independent of the stellar parameters and $G$. A similar argument holds for gas pressure-supported stars where $P\propto\rho T$, in which case one finds $\log(T_c)\propto\log(G)$. The true equation of state will be a mixture of the two. The effect of increasing $G$ at fixed mass is then to raise the central temperature at fixed central density, implying that tracks in the $\log(T_c)$--$\log(\rho_c)$ plane pass closer to the instability region. This is borne out in our numerical simulations, an example of which we plot in figure \ref{fig:Trho} where we plot the tracks at fixed mass for a $56{\rm M}_\odot$ star for three different values of $G$. The effect we describe here is clearly evident. The result of this is that the instability experienced is more violent, with two effects. First, more mass is lost during the pulses, resulting in lighter black holes, and, second, stars that would have undergone a PPISN now experience a PISN, which removes the heaviest black holes from the spectrum. The ultimate result of the altered stellar evolution is then to lower the edges (both upper and lower) of the BHMG.

A second effect of increasing $G$ is that the escape velocity
\begin{equation}
    v_{\rm esc}(r)=\sqrt{\frac{2GM(r)}{r}}
\end{equation}
is enhanced. This is relevant for both the pulsation and collapse phases. In particular, larger escape velocities implies that more mass is retained during the pulsations since less material becomes unbound, and that more material remains bound at core collapse. These effects compound to increase the final black hole mass, raising the edges of the mass gap. The highly dynamical and non-linear nature of the pulsations and core collapse preclude any analytic treatment of this effect. Furthermore, such a treatment is difficult numerically. This is because the pulsations begin after the helium burning phase when other effects of changing $G$ have already significantly altered the structure of the star. With no way of disentangling the separate effects, direct comparisons are difficult. Similarly, by the time stars of fixed initial mass have reached core collapse, their properties (including their masses) are different. 
\begin{figure*}[ht]
    \centering
    \includegraphics[width=0.85\textwidth]{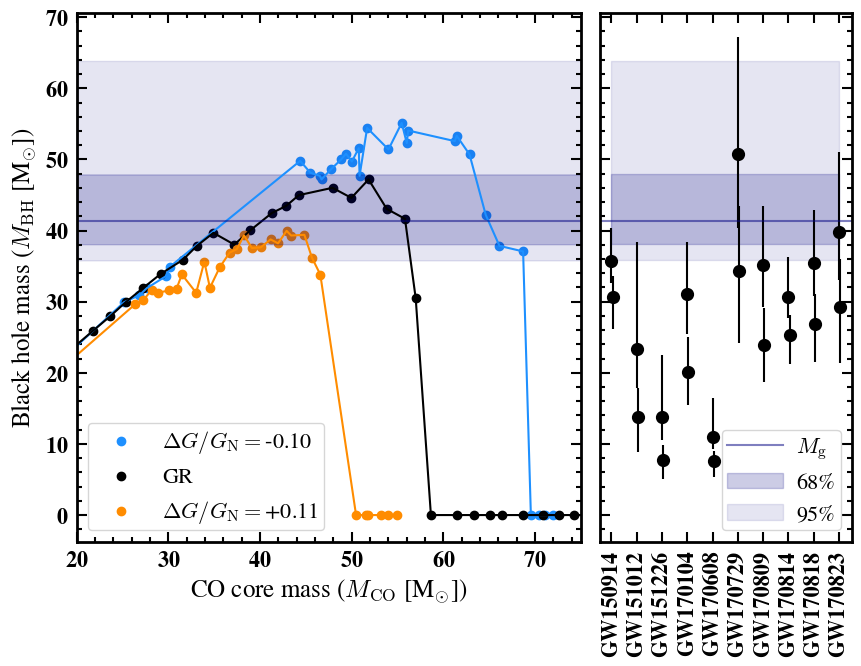}
    \caption{Black hole mass as a function of the star's carbon-oxygen (CO) core mass for values of $\Delta G/G_{\rm N}$ indicated in the figure. The metallicity of the progenitor stars was taken to be  $Z=10^{-5}$.  The right panel shows component masses of binary black hole mergers detected by the first and second LIGO/VIRGO observing runs \citep{Abbott2019-LIGO}. Error bars show 90\% confidence intervals on individual black hole masses. The navy horizontal line shows the median value for $M_{\rm g}$, and the shaded regions show 68\% (darker) and 95\% (lighter) confidence intervals on the lower edge of the black hole mass gap, computed from the ten detections as described in \S\ref{s.data}.
    }
    \label{fig:grids}
\end{figure*}

Finally, increasing $G$ reduces the lifetime of the helium burning phase. One can see this using scaling relations. The lifetime of helium burning is given by
\begin{equation}
    \tau_{\rm He}\sim \frac{M_{\rm He}}{L}
\end{equation}
where $L$ is the luminosity and $M_{\rm He}$ is the mass of helium in the core. Scaling arguments \cite{Davis:2011qf,Sakstein:2015oqa,Koyama:2015oma} predict that $L\propto GM$ for radiation pressure-supported stars and $L\propto G^4M^3$ { for gas pressure-supported stars } (the true equation of state is a mixture of the two), implying that increasing $G$ reduces the helium burning lifetime. The effects of this are two-fold. First, there is less time for mass to be lost due to stellar winds, which acts to raise the final black hole mass since more mass is retained at the onset of the pulsations, and, second, the ratio of \rC\ to \rO\ is increased. This is because there is less time for the ${}^{12}{\rm C}(\alpha,\,\gamma){}^{16}{\rm O}$ reaction, which converts \rC\,to \rO\,during core helium burning, to operate. Increasing the \rC\, to \rO\, ratio acts to quench the pulsations (either partially or fully depending on the star's mass and the ratio itself) \cite{Farmer2020-COreactionrate} resulting in less mass loss. The reasons for this are that oxygen is the fuel for the PPISN and PISN, and the convective motion of carbon being drawn from the shells into the core acts to suppress thermonuclear burning. The ultimate result for the reduced helium burning lifetime is then to allow heavier black holes to form, raising the edges of the mass gap. See \cite{Croon:2020ehi,Croon:2020oga} for more details on this. 

\begin{figure*}[ht]
    \centering
    \includegraphics[width=0.8\textwidth]{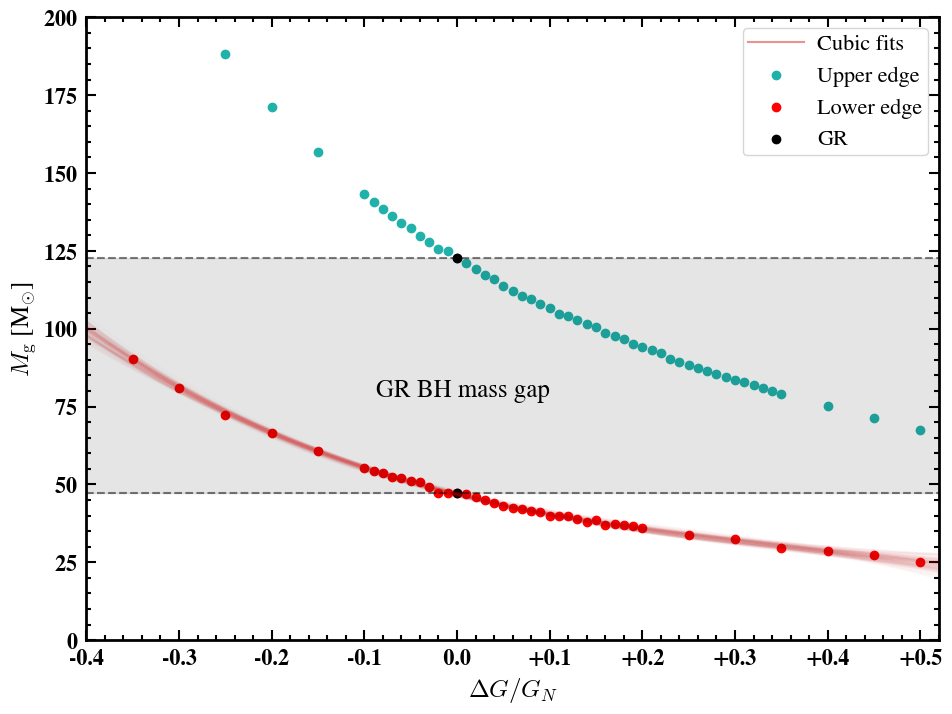}
    \caption{Upper and lower edges (shown by teal and red points respectively) of the black hole mass gap as a function of $\Delta G/G_{\rm N}$. Astrophysical black holes can form below the red points or above the teal points. The shaded region indicates the BHMG in GR and the black points indicate results for $\Delta G/G_{\rm N}=0$. The red  band show the results of fitting the lower edge measurements with a cubic polynomial, as described in \S\ref{s.data}; we adopt an uncertainty of $2{\rm M}_\odot$ for each point when performing this fit.
    }
    \label{fig:massgap}
\end{figure*}

We have run grids of \acro{MESA} simulations of the evolution of stars from the ZAHB to core collapse or PISN in intervals of $1{\rm M}_\odot$ for numerous values of $\Delta G/G_{\rm N}$ in the range $-0.35\le\Delta G/G_{\rm N}\le+0.50$ (the asymmetric range and frequency was determined by the requirements of the statistical analysis, see section \ref{s.data}). Our numerical investigations revealed that the most important effect of changing $G$ is the first one in our list above, namely that the altered stellar evolution results in more violent pulsations. This is exemplified in figure \ref{fig:MassvsTc} where we plot the mass vs central temperature (which is a proxy for time) of a star of initial mass $56{\rm M}_\odot$ evolving under GR and both reduced and enhanced values of $G$. Evidently, the differences in the amount of mass lost due to the alteration in the duration of the wind loss phase is negligible compared with those lost during the pulsations. The $\Delta G/G_{\rm N}=+0.07$ model exhibits two pulsations (steep vertical drops) in which $18.6{\rm M}_\odot$ of material is shed. The GR model exhibits three pulsations and sheds a total of $10{\rm M}_\odot$ of material. The $\Delta G/G_{\rm N}=-0.05$ model similarly experiences three pulsation but only sheds $6.6{\rm M}_\odot$ of material.  To exhibit the effects of this on the black hole mass function, we plot some representative grids in figure \ref{fig:grids}. Note that we plot black hole mass as a function of CO core mass rather than initial mass due to the tighter correlation \cite{Farmer2019-BHmassgap}. The effect of the more violent pulsations is evident from the lighter black holes formed when $\Delta G/G_{\rm N}>0$, and the effect of lighter stars undergoing PISN instead of PPISN is also evident. The ultimate effect of unscreening stars is then to lower the edge of the black hole mass gap (assuming $\Delta G/G_{\rm N}>0$, appropriate for the majority of screened modified gravity models).

In figure \ref{fig:massgap} we plot the black hole mass gap (both upper and lower edges) as a function of $\Delta G/G_{\rm N}$. The gray region in Fig.~\ref{fig:massgap} shows the location of the mass gap for $\Delta G/G_{\rm N} = 0$. From the figure, it is evident that increasing $\Delta G/G_{\rm N}$ lowers both edges.  The upper edge corresponds to the smallest mass progenitor where the PISN is quenched and so the effect of increasing $\Delta G/G_{\rm N}$ is to cause this to occur for lower mass objects. The shift in the lower edge is due to the effects discussed above, namely that the pair-instability is exacerbated, resulting in stronger pulsations, and that the PISN occurs in lighter objects, removing heavy black holes from the mass spectrum.

\section{Data Analysis and Constraints from Current Data}
\label{s.data}

We now use the LIGO/Virgo observations of BBHMs to place constraints on the modified gravity model described above.  Our analysis in this section is based largely on the statistical methodology developed in \cite{Fishbach2017-BigBlackHoles}, and we refer the interested reader to that work for more details.

\subsection{LIGO/Virgo Data}

Our analysis relies on BBHMs detected during the first and second observing runs of LIGO and Virgo, which ran from September 2015 to August 2017.  Ten BBHMs were detected at high significance in these runs, as shown in Fig.~\ref{fig:grids}.  We use the posterior samples on the detector frame component masses and redshifts made available at \url{https://dcc.ligo.org/LIGO-P1800370/public}, converting these to posterior samples on the source frame masses for the purposes of our analysis. 

\subsection{Methodology}

Observations of BBHMs by LIGO constrain the location of the BHMG \cite[e.g.][]{Fishbach2017-BigBlackHoles}.  Since all black hole detections from LIGO to date have masses significantly below the expected (in GR) upper edge of the BHMG, we focus our data analysis on the lower edge of the gap, which will yield tighter constraints on $\Delta G$ (although see the discussion of a recent BBH merger in \S\ref{sec:gw190521}). Armed with our knowledge of the location of the lower edge, $M_{\rm g}$, as a function of $\Delta G/G_{\rm N}$ (Fig.~\ref{fig:massgap}), constraints on $M_{\rm g}$ can be translated into constraints on $\Delta G/G_{\rm N}$, effectively constraining theories of modified gravity. { We remind the reader that we assume the waveform is unaltered by modified gravity i.e $\Delta G/G_{\rm N}=0$ for the black hole merger process. See \S\ref{s.waveform} for a discussion of this.}

One caveat of our analysis is our simplifying assumption that {\it all} black holes in the LIGO sample experience the effects of $\Delta G$.  This need not be the case if, for instance, some of the LIGO-detected black holes originated in screened environments.  For future analyses with large samples of well-localized BBHMs, it may be possible to first determine whether a black hole binary is in an unscreened environment using, e.g., the local distribution of galaxies \citep[e.g.][]{Desmond:2017ctk}.  Then, the methodology presented below could be applied only to the subset of systems that are believed to be unscreened. The present analysis should therefore be viewed as a proof of principle. Alternatively, the constraints derived below can be interpreted as limits on $G$ coming from extra degrees of freedom that satisfy no-hair theorems, independent of the motivation of screened modified gravity theories.

A second caveat of our analysis is that we consider the impact of modifying $G$ only on the location of the BHMG, and do not consider how changing $G$ may impact the black hole mass distribution in other ways, such as by changing the mass function of the stellar progenitors. As we discuss in more detail below, our analysis marginalizes over a parameter, $\alpha$, that characterizes the power law index of the underlying black hole mass function. As long as changing $G$ does not cause significant non-power law behavior in the black hole mass function, our analysis should yield correct results.  However, one could imagine a more sophisticated analysis that includes the impact of $\Delta G$ on all aspects of the black hole mass distribution, i.e. effectively making $\alpha$ a function of $\Delta G$. By including this additional information, it may be able to improve constraints relative to those presented here.

Given the LIGO/VIRGO-derived posterior samples on the primary and secondary (source frame) black hole masses, $m_1$ and $m_2$, we wish to obtain a posterior on $\Delta G$. Following \cite{Fishbach2017-BigBlackHoles}, we write the likelihood for the observations of the $i$th BBHM, $d_i$, as 
\begin{eqnarray}
\label{eq:likelihood}
P(d_i | \theta) \propto \frac{\langle P(m_1, m_2 | \theta) \rangle_{m_1 m_2}}{\beta(\theta)},
\end{eqnarray}
where $\theta$ represents the parameters specifying the mass distribution of the black holes (including $\Delta G$, which will modify the cutoff in this distribution), $P(m_1, m_2 | \theta)$ represents the probability of a BBH system with masses $m_1$ and $m_2$ given parameter values $\theta$, and the expectation value $\langle \ldots \rangle_{m_1 m_2}$ denotes an average over the posterior samples of $m_1$ and $m_2$ provided by LIGO for the $i$th event.  This average effectively integrates $P(m_1, m_2 | \theta)$ over the posterior distribution $P(m_1, m_2 | d_i)$, as shown in \cite{Fishbach2017-BigBlackHoles}.  The quantity $\beta$ is defined as
\begin{eqnarray}
\beta(\theta) \equiv \int P(m_1, m_2 | \theta) VT(m_1, m_2) dm_1 dm_2,
\end{eqnarray}
where $VT(m_1, m_2)$ is the population-averaged spacetime volume \cite{Fishbach2017-BigBlackHoles}.  If we are sensitive to a BBH merger of $m_1$ and $m_2$ over a small range of spacetime volume, then we are unlikely to measure $m_1$ and $m_2$ in the data, even if the mass distribution is such that these masses are common; $\beta(\theta)$ accounts for this fact.

We compute $VT(m_1, m_2)$ as detailed in \cite{Fishbach2017-BigBlackHoles}, using the expression for the optimal matched filter signal to noise from \cite{chen2014loudest, Finn_1996} and the estimated noise PSDs from \cite{noisepowerestimate}.  Since we view this analysis as a proof of principle, these noise power estimates should be sufficiently accurate for our purposes.

Information about $\Delta G$ enters via the quantity $P(m_1, m_2 | \theta)$, which effectively controls the mass distribution of the black holes. We begin by assuming that the probability distribution function for $m_1$ is given by a power law over $M_{\rm min} < m_1 < M_{\rm g}(\Delta G)$:
\begin{equation}
    P(m_1 | \theta) \propto m_1^{-\alpha} \mathcal{H}(M_{\rm g}(\Delta G)- m_1),
\end{equation}
where $\mathcal{H}$ is a Heaviside function, and we implicitly restrict $m_1 > M_{\rm min}$.  Assuming a uniform prior on the mass ratio for the binary pair, $q \equiv m_2/m_1 \leq 1$, over the allowed range of $q$, the joint mass distribution for $m_1$ and the mass of the smaller black hole, $m_2$, is given by 
\begin{equation}
    P(m_1, m_2 | \theta)  \propto \frac{m_1^{-\alpha}\mathcal{H}(M_{\rm g}(\Delta G)-m_1)}{{\rm min}(m_1, M_{\rm tot,max} - m_1) - M_{\rm min}},
\end{equation}
where $M_{\rm tot, max}$ is the maximum allowed value of $m_1 + m_2$, and again we implicitly restrict $m_1, m_2 > M_{\rm min}$.  The complete set of model parameters is therefore $\theta = \{\Delta G, \alpha, M_{\rm min}, M_{\rm tot,max} \}$.  As noted previously, the minimal black hole mass is thought to be roughly $5 \rm{M}_\odot$ \cite{Farr:2011}, and we set $M_{\rm min} = 5\,\rm{M}_\odot$ accordingly. Following \cite{Fishbach2017-BigBlackHoles}, we set $M_{\rm tot, max} = {\rm min}(2M_{\rm g}(\Delta G),100 \rm{M}_\odot)$.

Assuming uniform priors on $\Delta G$ and $\alpha$, the posterior $P(\Delta G, \alpha | d_i)$ is proportional to the likelihood in Eq.~\ref{eq:likelihood}.  We compute this posterior for each of the LIGO BBHM detections, and take the product across all events (under the assumption that they are independent) to compute a joint posterior:
\begin{equation}
    P(\Delta G, \alpha | \{ d_i \}) \propto \prod^N_{i=1} P(d_i | \Delta G, \alpha),
\end{equation}
where $N$ is the total number of BBHM events observed by LIGO.  Note that we can trivially replace $\Delta G$ by $M_{\rm g}$ in Eq.~\ref{eq:likelihood} as the parameter being varied in order to directly constrain $M_{\rm g}$ (i.e. for Fig.~\ref{fig:grids}).

As seen in Fig.~\ref{fig:massgap}, there is some scatter in the values of $M_{\rm g}$ computed by MESA.  Part of this uncertainty is due to the finite resolution of the grids we have used to compute $M_{\rm g}$ for each value of $\Delta G$; we refer to this scatter as numerical noise. We adopt an approximate model for the numerical noise, assuming that it is Gaussian, that it is uncorrelated between different masses, and that its amplitude is $\sigma = 2\rm{M}_\odot$.  Since the true relation between $M_{\rm g}$ and $\Delta G$ is likely to be a smooth function, we fit the data points shown in Fig.~\ref{fig:massgap} with a cubic polynomial, adopting the numerical noise model described previously.  The resultant errorband is shown in Fig.~\ref{fig:massgap} by the red lines.  As seen in the figure, this band fits the $M_{\rm g}$ data well, and allows for some reasonable level of uncertainty in the $M_{\rm g}(\Delta G)$ relation.  For each set of polynomial coefficients in the posterior chain resulting from the cubic fit, we compute the resultant posterior on $\Delta G$ and $\alpha$.  By summing these posteriors, we effectively marginalize over the uncertainty on the $M_{\rm g}(\Delta G)$ relation from numerical noise.

Note that the numerical noise is not the only source of uncertainty in the $M_{\rm g}(\Delta G)$ relation.  Other sources of uncertainty, such as in various nuclear reaction rates, can also lead to uncertainty in this relation. In the interest of simplicity, we ignore these additional sources of uncertainty in this first analysis.  

\begin{figure}
    \centering
    \includegraphics[width=3.25in]{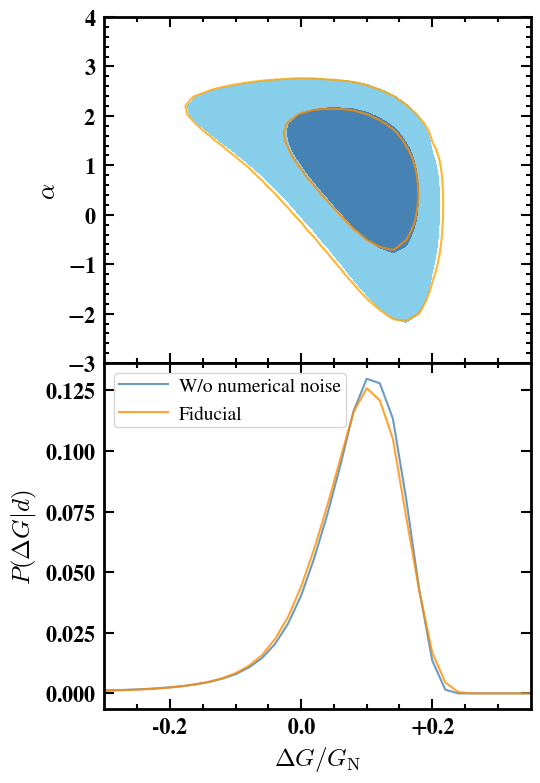}
    \caption{{\it Top panel}: the joint posterior on $\Delta G/G_{\rm N}$ and the mass function power law index $\alpha$. Blue contours show the posterior without including the impact of numerical noise in the $M_{\rm g}$ vs. $\Delta G/G_{\rm N}$ relation, while the orange curves do include this source of uncertainty.  Inner and outer contours enclose 68\% and 95\% of the posterior mass, respectively.  {\it Bottom panel}: constraints on $\Delta G/G_{\rm N}$ after marginalizing over $\alpha$. We find $\Delta G/G_{\rm N} = 0.1^{+0.04}_{-0.1}$ at 68\% confidence.
    }
    \label{fig:constraints}
\end{figure}

\subsection{Constraints on $\Delta G$ From Current Data}

The joint posterior on $\Delta G$ and $\alpha$ is shown in the top panel of Fig.~\ref{fig:constraints}. The blue contours indicate the posterior in the absence of the numerical noise described above, while the orange curves include this source of uncertainty.  We find that the scatter about the $M_{\rm g}(\Delta G)$ relation caused by numerical noise has an essentially negligible impact on our posteriors.  

Our constraints on the $\Delta G/G_{\rm N}$ and $\alpha$ parameters are somewhat degenerate, with larger $\alpha$ preferring lower $\Delta G$.  As $\alpha$ is increased, the mass distribution of BHs falls off more steeply; this effect can be offset by pushing $M_{\rm g}$ to higher values by decreasing $\Delta G$.  These effects are not perfectly degenerate, though, as $\alpha$ changes the distribution of BH masses near the gap, while $\Delta G$ in our model only changes its location. As noted before, since we have ignored the impact of modified gravity on the mass distribution of stellar progenitors in our analysis, the constraints on $\alpha$ should be interpreted with caution.  

Marginalizing over $\alpha$ yields the constraints on $\Delta G$ shown in the bottom panel of Fig.~\ref{fig:constraints}.  We find $\Delta G/G_{\rm N} = 0.1^{+0.04}_{-0.1}$ at 68\% confidence.  The data prefer a somewhat positive $\Delta G$, but are consistent with $\Delta G/G_{\rm N} = 0$ (i.e. GR) at 68\% confidence.  Again, the impact of numerical noise is minimal.  We also repeat our analysis adopting $M_{\rm min} = 2 \rm M_{\odot}$ rather than $M_{\rm min} = 5\rm M_{\odot}$; this shifts the posterior very slightly to higher $\Delta G$, but the shift is almost completely negligible relative to the errorbars.

\subsection{Comparison with Existing Bounds}

The analysis above assumed that all ten LIGO/Virgo BBM events were unscreened. Extending this analysis to theories with screening mechanisms requires one to account for either the time-dependence of $G$ or environmental screening that we discuss in section \ref{s.concs}. Despite this, it is instructive to compare the bound we derived above with others already in the literature to determine if or when our test could become competitive with these.

\subsubsection{Bounds on Screened Modified Gravity}

{For theories such as dark matter--baryon screening, one can obtain the bounds $\Delta G/G_{\rm {N}}<0.05$ from a comparison of Cepheid and tip of the red giant branch distances to unscreened galaxies \cite{Desmond2019-resolveHT,Desmond2020-Screened} so it is likely that competitive constraints will be achieved only after LIGO/Virgo's sensitivity is upgraded and the number of detections has increased sufficiently. Modified gravity theories that screen via the thin-shell effect e.g. chameleon models typically predict $0.1\le\Delta G/G_{\rm N}\le1$. Constraints on $\Delta G/G_{\rm N}$ apply as a function of a second parameter that determines the ability of an object to self screen (denoted $\chi_0$ or $f_{R0}$ in the literature). Population-III stars have Newtonian potentials of order $GM/Rc^2\sim 10^{-8}$ so are some of the most unscreened objects in the universe. It is then likely that the lower edge of the BHMG could form a novel and powerful probe of these theories, if a suitable observational strategy that accounts for environmental screening can be developed (see section \ref{s.obs_strat} for a discussion of this).} {For theories that do not include environmental screening, such as DHOST and beyond Horndeski, the Sun and other Milky Way objects provide strong constraints on $\Delta G/G_{\rm N}$ \cite{Sakstein:2015aac,Sakstein:2015zoa,Saltas:2019ius,Crisostomi:2019yfo}. In these theories, $\Delta G/G_{\rm N}$ is time-dependent, so it is possible that our black hole mass gap test, which can reach luminosity distances of order Gpc, could complement these low redshift bounds by testing this feature. }

{Our results indicate that the upper edge of the BHMG is a promising novel probe of modified gravity. Reference \cite{Ezquiaga2020-jumpthegap} have demonstrated that black holes from the upper edge may be detectable by LIGO/Virgo once they are upgraded to `A+' sensitivity. A striking prediction of modified gravity theories is that these upper edge objects would exist with lower masses, making them more readily detectable. For large enough values of $\Delta G/G_{\rm N}\gtrsim+0.3$, black holes with masses $M_{\rm BH}\lsim85{\rm M}_\odot$ are predicted. These lie squarely in the mass gap predicted by GR and are detectable with LIGO/Virgo's current sensitivity. Chameleon theories, and in particular $f(R)$ theories, predict $\Delta G/G_{\rm N}=1/3$ so this effect represents a promising probe of such models.}


\subsubsection{Bounds on Strong Equivalence Principle Violations}

The bound obtained in this analysis is strictly a bound on the strong equivalence principle between black holes and baryonic matter. Similar tests are few and far between, with the most notable being the prediction of offset supermassive black holes from the centers of galaxies due to their insensitivity to fifth forces implied by the no-hair theorem \cite{Hui:2012jb}. Reference \cite{Asvathaman:2015nna} found a bound on the E\"{o}tv\"{o}s parameter $\eta<0.68$ for black holes and baryons using this effect applied to M87. Theories such as galileons that screen using the Vainshtein mechanism are difficult to test using conventional methods due to their high screening efficiency. For this reason, strong equivalence violation tests are the most constraining for these theories \cite{Sakstein:2017bws}.

\subsection{Improvements with Future Data Sets}

With the analysis described above, we have achieved a roughly 7\% constraint on $\Delta G$ using ten BBHM detections from the first two runs of LIGO and Virgo.  Future data from LIGO, Virgo, and KAGRA \cite{Kagra} will significantly expand the sample size of BBHM events.  Interestingly, the constraints on a parameter representing the maximum value in a population of observed events (such as $M_{\rm g}$) can improve faster than the usual $1/\sqrt{N}$, where $N$ is the number of observed events (see Appendix \S\ref{app:toymodel} for a toy model that illustrates this fact).  Consequently, we expect constraints on $M_{\rm g}$ and thus $\Delta G$ to improve significantly in the future.  O3 observations from LIGO/Virgo are expected to increase the number of BBHM events by roughly a factor of five, leading to a decrease in the $\Delta G$ uncertainty by a corresponding factor of five, since the dependence of $M_{\rm g}$ on $\Delta G$ in the region of interest is roughly linear. Improved constraints on $\Delta G$ can also be achieved by including more information in the analysis. For instance, one could include the impact of $\Delta G$ on the entire distribution of BBH masses, rather than its impact on only the location of the mass gap. We postpone such an analysis for future work.

As suggested by the right panel of Fig.~\ref{fig:grids}, current uncertainties on the BBH component masses significantly degrade our constraints on the location of the mass gap, and thus $\Delta G$. Additional detections with reduced error bars --- for example from systems with highly asymmetric mass ratios, which help to break some degeneracies --- would allow for tighter bounds.


\section{Discussion and Conclusions}
\label{s.concs}

\subsection{Observational Signatures and Detection Strategies}
\label{s.obs_strat}

In the previous section we demonstrated the constraining power of the BHMG by obtaining a bound on the ratio of the gravitational constant experienced by black holes and baryonic matter, a novel test of the strong equivalence principle, which is generically violated in theories beyond general relativity. Implicit in this analysis was the assumption that the change in this due to modified gravity is universal, implying the absence of a screening mechanism. Theories without screening mechanisms are best probed using other means e.g. laboratory tests or test of post-Newtonian gravity so our primary focus is on those that do include such mechanisms. In this case, the environmental dependence of the fifth force strength implies that the change in the location of the BHMG is not universal but instead that there exist two separate populations of astrophysical black holes: those that formed in screened galaxies where $\Delta G/G_{\rm N}=0$ and those that formed in unscreened galaxies where $\Delta G/G_{\rm N}>0$ and galaxy-dependent. The former are by far the more numerous. In this section we discuss several potential detection strategies for this smaller population, postponing the development of such pipelines for future work.

\begin{itemize}
    \item {\bf Black hole population statistics:} With the large number of detections expected in the coming years, it will not only be possible to measure the position of the lower edge with sub-${\rm M}_\odot$ precision \cite{Fishbach:2019bbm} but also to perform detailed population studies \cite{LIGOScientific:2018jsj}. The expected population can be computed using the initial mass function and by accounting for other astrophysical effects \cite{vanSon:2020zbk}. The mass gap for the black hole population in unscreened galaxies begins at lighter masses, implying a lack of heavy black holes. A sparser population of observed black holes around the lower edge of the BHMG predicted by GR could then form a potential observation probe. 
    
    \item {\bf Detections in the mass gap:} For sufficiently large values of $\Delta G/G_{\rm N}\gtrsim 0.2$ (these values are typical of chameleon and $f(R)$ theories) the black holes formed at the upper edge of the modified gravity mass gap lie squarely in the mass gap predicted by GR. Similarly, for $\Delta G/G_{\rm N}\lsim-0.25$. Having masses in the range $60{\rm M}_\odot\le M\le90{\rm M}_\odot$, these objects are detectable by LIGO/Virgo. The detection of these objects could then constitute a novel probe of screened modified gravity. There are some potential backgrounds for the detections of such black holes. Black holes that formed from previous black hole-black hole mergers \cite{Mangiagli:2019sxg} or that accrete gas from proto-globular clusters \cite{Roupas:2018cvb, Roupas:2019dgx} could have masses that place them within the GR BHMG. The rate of such mergers is expected to be significantly smaller than the rate of black hole formation from the core collapse of population-III stars \cite{vanSon:2020zbk}. Similarly, it is possible to form black holes in the mass gap from population-I or population-II stars due to stellar process that significantly inhibit wind loss. \cite{Limongi:2018qgr,Belczynski:2019fed}.  Modeling and characterizing these populations e.g. \cite{Flitter:2020bky} would aid in reducing these backgrounds and enabling constraints on any additional events due to screened modified gravity. Recently, LIGO/Virgo announced the detection of a binary black hole merger with both components in the GR mass gap: GW190521  \cite{PhysRevLett.125.101102,Abbott_2020}. We comment on this below. See \cite{Abbott_2020} for a discussion of possible standard model formation mechanisms for the heavy black holes inferred from this event.
    
    \item {\bf Time-variation of $G$:} There has been a recent interest in modified gravity theories where there is a strong time-variation of $G$. In particular, theories where $G$ increases in the past may provide a partial resolution of the Hubble tension \cite{Ballesteros:2020sik,Braglia:2020iik}. This would imply that there is a redshift-dependence of the location of the BHMG, with the lower edge shifting to smaller values in the past and the upper edge potentially being visible. This could be searched for by splitting the LIGO/Virgo observations into different luminosity distance bins and looking for such a variation. The current number of data points are too few and the current error bars are too large to allow for a meaningful attempt at this at the present time. A sufficient number of events with optical counterparts may allow for a similar test in redshift space. {Some screened theories such as DHOST and beyond Horndeski also include a time-dependence of $G$, and it is possible that they too could be constrained using this method.}
    
    \item {\bf Localization of specific events:} It is possible to localize the origin of binary black hole mergers. If the event is accompanied by an electromagnetic counterpart then this is possible using follow up observations to determine the position of the optical source but even if no such counterpart is observed, one can correlate the LIGO/Virgo detection region with galaxy catalogs to find candidate host galaxies \cite{Nissanke:2011ax,Aasi:2013wya,Oguri:2016dgk,Bartos:2017ggb,Calore:2020bpd}. The screening status of the host galaxy can then be determined using screening maps \cite{Cabre:2012tq,Desmond:2017ctk,Shao:2019wit,Desmond:2020gzn}, which determine whether an individual galaxy is screened or not and provide the (environment-dependent) value of $\Delta G/G_{\rm N}$ as a function of the theory parameters. Localizing events containing black holes in the GR mass gap can then directly constrain $\Delta G/G_{\rm N}$ and the other model parameters where present. LIGO/Virgo's ability to localize specific events will dramatically improve once additional detectors come online \cite{Pankow:2018phc}. 
    
\end{itemize}

\subsection{Impact of Modified Gravity on the Waveform}
\label{s.waveform}

Throughout the data analysis, and the discussion of observational strategies above, we assumed that modified gravity has no effect on the waveform observed by LIGO/Virgo. This assumption is well motivated by the no-hair theorems. Even though their progenitor stars are sensitive to the modifications of gravity through the effects of fifth forces, the black holes themselves are not due to absence of any hair (see \cite{Hui:2012jb} for a discussion of this). One potential caveat to this is if a large amount of radiation in the form of additional polarizations of the graviton are emitted during the merger, especially in the form of monopole or dipole modes. This is not expected to be the case in the theories of interest since the same non-linearities responsible for the screening mechanisms suppress the power emitted in these modes \cite{deRham:2012fg,Dar:2018dra}. Another possibility is that the signal could be produced as in GR but altered by new effects on cosmological scales such as damping from a running of the Planck mass \cite{Mastrogiovanni:2020tto,Baker:2020apq}. In light of this, the results of our data analysis are not a measurement of Newton's constant because changing this for all objects would also impact the waveform,  but more of a test of the strong equivalence principle by constraining deviations between the strength of gravity felt by black holes and baryonic matter. In light of this discussion, we emphasize that the application of our results to constraining modified gravity theories must be considered on a theory-dependent basis.

\subsection{GW190521}
\label{sec:gw190521}

As this work was being prepared, the LIGO/Virgo collaboration announced the discovery of GW190521 \cite{PhysRevLett.125.101102,Abbott_2020}. This was identified as a binary black hole merger with components $m_1 =85^{+21}_{-14}{\rm M}_\odot$ and $m_2=66^{+17}_{-18}{\rm M}_\odot$ (although see \cite{Fishbach:2020qag} for an alternative interpretation where one component is a lower edge black hole and the other is an upper edge black hole). Both of these lie directly inside the canonical black hole mass gap predicted by general relativity. In light of the results of our study, these objects could have formed if their host galaxy was unscreened provided { $\Delta G/G_{\rm N}=0.3_{-0.2}^{+0.15}$ or $\Delta G/G_{\rm N}=-0.3_{-0.1}^{+0.05}$. The lower limit in the former case corresponds to the upper limit on $m_1$ and conversely for the upper limit. In the latter case, the upper limit corresponds to the lower limit on $m_1$ and conversely for the lower limit.} We consider the former scenario more likely on theoretical grounds. There is no definitive way to falsify this hypothesis. An accompanying electromagnetic counterpart could have allowed for a localization of the host galaxy, in which case its screening status could be determined. A candidate event was identified by the Zwicky Transient Facility in the form of a flare \cite{Graham:2020gwr}. Such events are not typically associated with merging black holes, and there was no prompt counterpart reported by any other facilities. Whether this event was indeed associated with GW190521 remains unclear. { Either way, the screening maps to do not currently extend to the distances associated with this event ($\sim 5$ Gpc), so we are unable to determine the screening status of the associated host object.} Reference \cite{Sakstein:2020axg} examines possible beyond the standard model explanations of GW190521, including modified gravity.  

\subsection{Conclusions}

In this work we have initiated the study of the effects of modified gravity on the location of the black hole mass gap. The theories motivating this are those that are highly relevant to dark energy and the Hubble tension, namely those with screening mechanisms. A generic feature of these theories is the presence of new or fifth forces that arise from the coupling of light gravitational degrees of freedom (typically scalar fields) to matter. Screening mechanisms hide these in the solar system, thereby allowing consistency with laboratory and solar system tests of gravity, but they can emerge on cosmological scales to drive the acceleration of the cosmic expansion. 

On intermediate scales, unscreened galaxies and their constituents exhibit novel phenomenologies, making them prime laboratories for testing these theories. The effects on the BHMG that we have studied in this arise because the population-III progenitors of the binary black holes observed by LIGO/Virgo are unscreened in these galaxies and experience the fifth force. This acts to alter the structure and evolution of these objects because the conditions required to maintain hydrostatic equilibrium are modified. In theories where the strength of gravity is enhanced in unscreened environments (this is the majority of theories of interest), the effects of the fifth force are to raise the star's central temperature at fixed initial mass and central density. This results in a larger number of electron-positron pairs being produced thermally, exacerbating the pair-instability. There are two consequences of this. At fixed mass, the pulsations are more violent and more mass is shed when they are active. Similarly, stars with masses that would have implied a pulsational pair-instability now end their lives in a pair instability supernova instead. The ultimate effect on the black hole mass distribution is that heavier black holes disappear from the spectrum, resulting in a shift in the location of the lower edge of the mass gap towards lighter masses. Similarly, the quenching of the pair-instability due to the photodisintegration of heavy elements occurs in lighter mass objects, resulting in the upper edge moving towards lighter masses. Interestingly, for values of $\Delta G/G_{\rm N}$ relevant for the chameleon screening mechanism (and similar) we found that the upper edge black holes would lie in the mass gap predicted by GR. These objects have masses $60{\rm M}_\odot<M<90{\rm M}_\odot$, and are observable by LIGO/Virgo.  They may even explain the recent GW190521 binary black hole merger event, for which both merging black holes have inferred masses in this range.

As a proof of principle to demonstrate the constraining power of the mass gap, we performed a statistical analysis of the first ten LIGO/Virgo binary black hole merger events to obtain a measurement of the value of Newton's constant. We find $\Delta G/G_{\rm N} = 0.1^{+0.04}_{-0.1}$ at 68\% confidence. {The theories probed by this measurement are those without screening mechanisms}. The true power of our results lie in their ability to constrain theories that do utilize screening mechanisms, and for which solar system tests are unconstraining. This requires the development of dedicated detection strategies that account for the uncertainties introduced by environmental screening and possibly modifications of the waveform. We have discussed several possibilities at length, and proposed several methods by which they could be executed. We intend to develop this program in future work. 

\acknowledgements

We are grateful to Robert Farmer, Pablo Marchant, and the MESA community for answering our many questions about MESA. We would like to thank Djuna Croon, Harry Desmond, Samuel D. McDermott, Maya Fischbach, and Robert Jedicke for enlightening discussions. MCS acknowledges support from the Research Experience for Undergraduates program at the Institute for Astronomy, University of Hawai'i-Manoa funded through NSF grant 6104374.

\bibliography{references}

\appendix
\section{Toy Model Illustrating Improvement in Constraints on the Maximum of a Distribution}
\label{app:toymodel}

In this appendix we show that for some sample of observed events, $\{ x_1, x_2 \ldots x_N \}$, the constraints on the  inferred population maximum, $x_{\rm max}$, can improve faster than the usual $1/\sqrt{N}$, where $N$ is the number of observed events.

For simplicity, we assume that the population is uniformly distributed between 0 and $x_{\rm max}$, i.e.
\begin{eqnarray}
x_i \sim U(0,x_{\rm max}),
\end{eqnarray}
where $U(a,b)$ represent the uniform distribution between $a$ and $b$.  The likelihood for an individual event is
\begin{eqnarray}
P(x_i | x_{\rm max}) \propto 
\begin{cases}
1/x_{\rm max} \textrm{ for $x_i < x_{\rm max}$} \\
0 \textrm{ otherwise}.
\end{cases}
\end{eqnarray}
Adopting a flat prior on $x_{\rm max}$ and assuming the $x_i$ are independent, the total posterior is 
\begin{eqnarray}
P(x_{\rm max} | \{ x_i\}) = 
\begin{cases}
\frac{A}{x_{\rm max}^N} \textrm{ for $x_{\rm max} > \mathrm{max}[x_i]$} \\
0 \textrm{ for $x_{\rm max} < \mathrm{max}[x_i]$},
\end{cases}
\end{eqnarray}
where $A$ normalizes the distribution, and its value is given by
\begin{eqnarray}
A = \int_{\rm{max}[x_i]}^{\infty} dx_{\rm max} \frac{1}{x_{\rm max}^N} = \frac{(\mathrm{max}[x_i])^{1-N}}{N-1}.
\end{eqnarray}

The first moment of the total posterior is 
\begin{eqnarray}
\langle x_{\rm max} \rangle = (1/A) \int_{\rm{max}[x_i]}^{\infty} dx_{\rm max} \frac{x_{\rm max}}{x_{\rm max}^N} = \frac{\mathrm{max}[x_i](N-1)}{(N-2)}, \nonumber \\
\end{eqnarray}
and the second moment is
\begin{eqnarray}
\langle x^2_{\rm max} \rangle = (1/A) \int_{\rm{max}[x_i]}^{\infty} dx_{\rm max} \frac{x_{\rm max}^2}{x_{\rm max}^N} = \frac{(\mathrm{max}[x_i])^2(N-1)}{(N-3)}. \nonumber \\
\end{eqnarray}
The variance is then given by
\begin{eqnarray}
\mathrm{var}(x_{\rm max}) &=& (\mathrm{max}[x_i])^2 \frac{(N-1)}{(N-3)(N-2)^2} \\
&\sim& \frac{(\mathrm{max}[x_i])^2}{N^2},
\end{eqnarray}
and we see that the uncertainty on $x_{\rm max}$, $\sigma(x_{\rm max}) \equiv \sqrt{{\rm var}(x_{\rm  max})}$, will go scale as $1/N$ rather than the usual $1/\sqrt{N}$.

\end{document}

%% file: universalnewcommands.tex
\newcommand{\lsim}{\lesssim}

\newcommand{\acro}[1]{\textsc{\MakeLowercase{#1}}} 

\newcommand{\beq}{\begin{equation}}
\newcommand{\eeq}{\end{equation}}
\newcommand{\bea}{\begin{eqnarray}}
\newcommand{\eea}{\end{eqnarray}}

\newcommand{\refjnl}[1]{{\rm#1}}

\def\apj{\refjnl{ApJ}}                 
\def\apjs{\refjnl{ApJS}}               
\def\apss{\refjnl{Ap\&SS}}             
\def\aap{\refjnl{A\&A}}                

\def\prd{\refjnl{Phys.~Rev.~D}}        
